\pdfoutput=1

\documentclass[11pt]{article}

\usepackage{ACL2023}

\usepackage{times}
\usepackage{latexsym}

\usepackage[T1]{fontenc}

\usepackage[utf8]{inputenc}

\usepackage{microtype}

\usepackage{inconsolata}

\usepackage{amsmath,amsfonts}
\usepackage{algorithmic}
\usepackage{graphicx}
\usepackage{textcomp}
\usepackage{xcolor}
\usepackage{algorithm}
\usepackage{multirow}
\usepackage{booktabs}  
\usepackage{bm}
\usepackage{arydshln}

\title{CodePAD: Sequence-based Code Generation with Pushdown Automaton}

\author{Yihong Dong, Xue Jiang, Yuchen Liu, Ge Li\thanks{\quad Corresponding Author} , Zhi Jin\footnotemark[1]\\
Key Lab of High Confidence Software Technology (PKU), Ministry of Education \\
School of Computer Science, Peking University, China\\
\texttt{\{dongyh, jiangxue\}@stu.pku.edu.cn, \{lige, zhijin\}@pku.edu.cn} \\}

\begin{document} 
\maketitle
\begin{abstract}
    In the process of code generation, it is essential to guarantee the generated code satisfies grammar constraints of programming language (PL). However, neglecting grammar constraints is a fatal drawback of commonly used sequence-based code generation. In this paper, we devise a pushdown automaton (PDA)-based methodology to address this problem, exploiting the principle that PL is a subset of PDA recognizable language and code accepted by PDA is grammatical. Specifically, we construct a PDA module and design an algorithm to constrain the generation of sequence-based models to ensure grammatical correctness. Guided by this methodology, we further propose CodePAD, a sequence-based code generation framework equipped with a PDA module, to integrate the deduction of PDA into deep learning. Additionally, this framework can leverage states of PDA deduction (including state representation, state prediction task, and joint prediction with state) to assist models in learning PDA deduction. To comprehensively evaluate CodePAD, we construct a PDA for Python and conduct extensive experiments on four public benchmark datasets. CodePAD can leverage existing sequence-based models, and we show that it can achieve 100\% grammatical correctness percentage on these benchmark datasets. Thus, it relatively improve 17\% CodeBLEU on CONALA, 8\% EM on DJANGO, and 15\% CodeBLEU on JUICE-10K compared to base models. In addition, our method significantly enhances pre-trained models, e.g., CodeBLEU of CodeGen-350M improvement from 3.21 to 21.54 on MBPP in zero-shot setting. 
\end{abstract}

\section{Introduction}
Code generation is a hot research spot in the field of natural language processing (NLP) and software engineering \cite{Ling16, TranX, Wei2019, APT, IndustryCodeGeneration}, which aims to help facilitate software development and revolutionize user programming. 
A well-designed code generation approach requires consideration of grammar in addition to semantics, as semantics ensures the functionality of code meets developer's intention, while satisfying grammar is a basic requirement for programming.

Sequence-based approaches are the most commonly used in code generation \cite{Raychev14, JiaL16, Cao19, CodeT5, Incoder, Codegen}, which has the following three advantages: 1) High efficiency. Sequence-based approach directly generates a token sequence of code along the order of human writing in one pass. 2) Convenient operation. It can obtain partially generated code with ease even if the result is incomplete or the generation is not finished, thus can be adopted for various code generation scenarios \footnote{For example, code completion, a popular variant of code generation in practice, which completes code with contexts.}. 
3) Easy data accessibility. Sequence structured data is the most general form that can be obtained without much effort compared to other structured data. 
Thus, pre-trained methods with big data-driven usually employ sequence-based approaches \cite{CodeT5, AhmadCRC21PLBART, UniXcoder, Incoder, Codegen}. However, most sequence-based approaches generate codes with no guarantee of grammatical correctness (GC), which hinders the usability of generated codes.

According to automata theory, we find that PDA is suitable for PL grammar and has the following properties for dealing with non-grammatical problems. The first property is a valid terminal symbol set for next token prediction, and the second is an accept state set for recognizing completed code. In a practical code generation process, what we really care about are two typical grammatical error: terminal symbol mismatch (TSM) error and end with non-accept state (ENS) error, where ENS error can be tolerable until the end of code generation process. As a result, satisfying grammar constraints during code generation require the above two properties, and constraining code generation based on PDA to guarantee GC is a matter of course.  


\begin{figure}[t]
	\centering
	\includegraphics[width=0.5\textwidth]{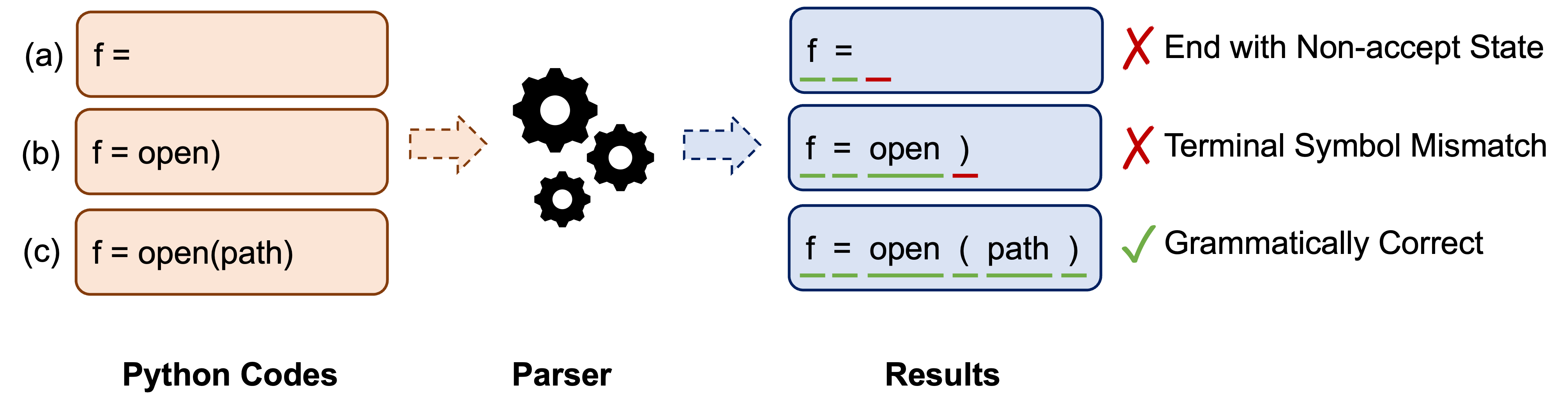}
	\caption{Examples of a parser parsing Python code.}
	\label{parser}
\end{figure}

It is inspired by the operating principle of programming language parsers in practice. Fig. \ref{parser} demonstrates some examples of a parser parsing python code, which contains two specific grammatical error codes (a) and (b), and a grammatically correct code (c). Case (a) indicates ENS error, and case (b) indicates TSM error. All cases are recognized by a parser according to grammar. Therefore, a straightforward way is applying a parser to constrain sequence-based code generation, but it has the following two problems. First, sequence-based models need to recognize grammatical errors during code generation process. However, the vast majority of codes generated during this situation are incomplete, i.e., parsing these codes will produce ENS errors like case (a). Second, a parser can only recognize TSM errors like case (b) if terminal symbol is fed to the parser. It implies that sequence-based models need to attempt each generated code token in the candidate set one by one, which is strenuous \footnote{For example, there are only 83 syntax-strings and 10 token-types in Python, and for a 10k-length vocabulary, more than 99\% of tokens belong to three token-types NAME, STRING, and NUMBER. If models want to generate these types of tokens based on semantics but are prevented by grammar, the cost of attempts could be overwhelming.}. In short, although code can be grammatically bounded by a parser, it is still challenging to work directly with a parser on sequence-based code generation. 


In this paper, we propose CodePAD (a \underline{Code} generation framework based on \underline{P}ushdown \underline{A}utomaton mo\underline{D}ule) to ensure GC for sequence-based models. PDA module facilitates CodePAD to generate bounded next prediction in a valid set and end with an accept state, utilizing the principle that PDA can recognize grammatical codes. CodePAD not only ensures GC but also maintains the advantages of commonly used sequence-based code generation. We conduct a series of experiments on four public benchmark datasets. Extensive experimental results and analyses verify the effectiveness and generality of CodePAD. The main contribution of this paper can be summarized as follows:
\begin{itemize}
	\item We devise a PDA-based methodology to guarantee grammatical correctness for code generation, which consists of a PDA module and an algorithm to simulate the deduction of PDA. 

    \item We propose CodePAD, a novel sequence-based code generation framework incorporating the PDA module. In addition to using the PDA-based methodology, CodePAD leverages PDA state through state representation, state prediction task, and joint prediction with state, which helps learn PDA deduction.
	
	\item CodePAD significantly enhances the performance of base models. Even in zero-shot setting, pre-trained models still show remarkable improvements.
	
\end{itemize}

\section{Motivation Example}
\begin{figure}[ht!]
	\centering
	\includegraphics[width=0.5\textwidth]{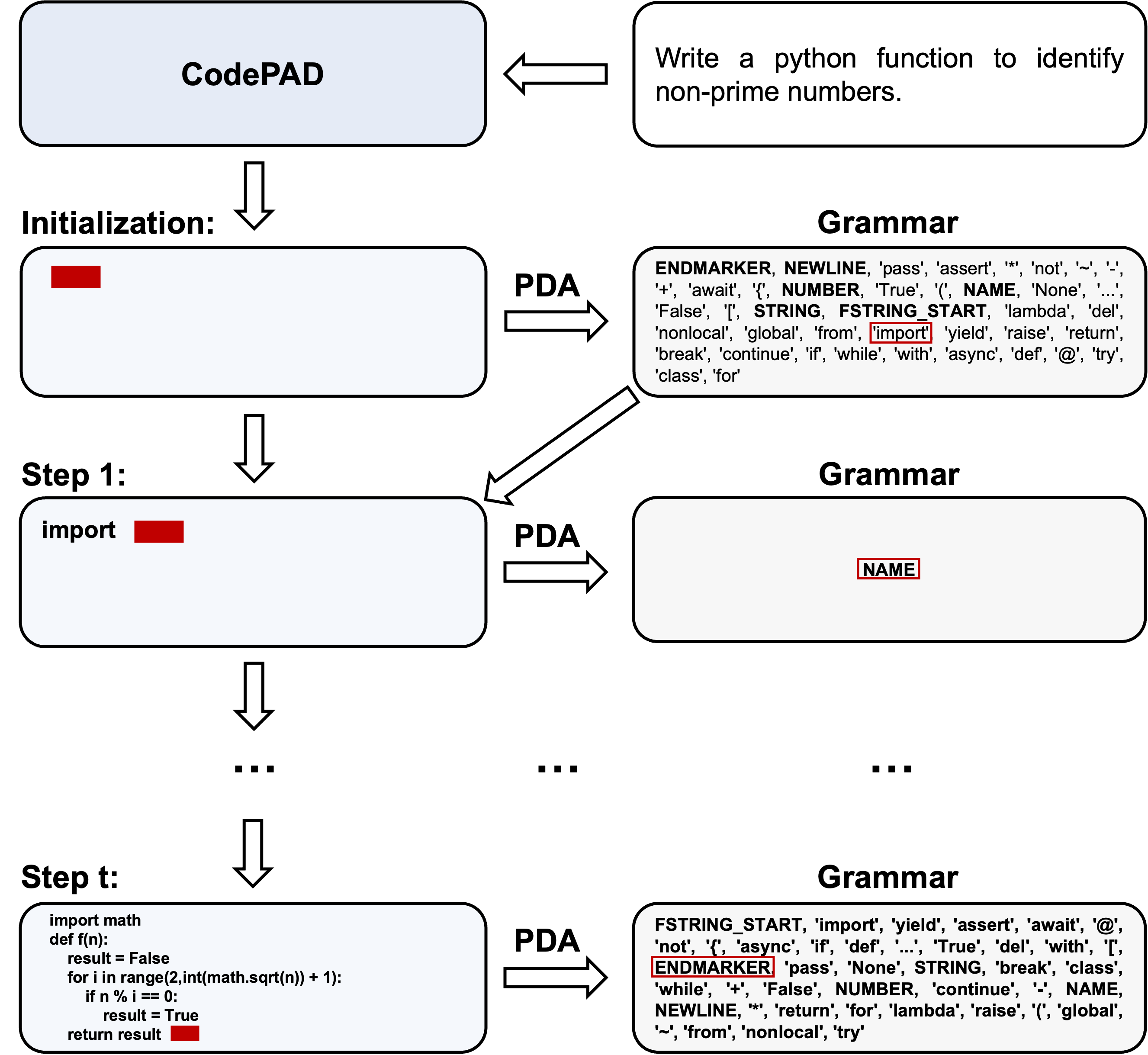}
	\caption{Motivation Example.}
	\label{Motivation Example}
\end{figure}
An example of code generation constrained by PDA module is shown in Fig. \ref{Motivation Example}. Given an NL, CodePAD needs to generate code that satisfies its intended functionality with the guarantee of GC. PDA module gives a valid set of terminal symbols (i.e., candidate syntax-strings and token-types) depending on grammar constraints, to which the generated token of CodePAD should belong. 
Specifically, in the beginning, CodePAD wants to generate first token from the vocabulary of a model according to NL. Based on production rules, PDA module shrinks the candidate set from entire vocabulary to 35 syntax-strings and 6 token-types, where each candidate token is constrained by grammar. As a result, at step 1, CodePAD picks a token (i.e., `import') from the candidate set according to semantics. At each subsequent generation step, CodePAD predicts a token in the same manner, which satisfies grammar constraints.
Until the end of generation steps, CodePAD outputs the token of ENDMARKER token-type (e.g., </s>), which ends with an accept state. Thus, CodePAD generates a grammatical code with the help of PDA module.


In the way of the above example, we first introduce PDA into deep learning (DL), which drastically shrinks the candidate set at each step of generation and reduces the difficulty for the model to select tokens from it, thus ensuring GC and improving the quality of code generation.

\section{Related Work}
\label{Related Work}
\subsection{DL-based Code Generation}
In recent years, two kinds of prevailing approaches have been used for DL-based code generation. 

\subsubsection{Sequence-based Approaches} 
\cite{Ling16} considered code generation task as a conditional text generation task and adopted a sequence-based model to address it. Some later work \cite{JiaL16, Cao19} followed this generation approach. For example, \cite{Wei2019} proposed a dual training framework to train both code generation and representation tasks simultaneously. CodeT5 \cite{CodeT5}, UniXcoder \cite{UniXcoder},  Codegen \cite{Codegen}, and InCoder \cite{Incoder} applied pre-trained models to code generation task. In addition, \cite{Wang22} considered code compilability as a training objective based on pre-trained models, but it still fails to guarantee GC. 

Although sequence-based approaches are most commonly used, they usually cannot guarantee GC.

\subsubsection{Tree-based Approaches}
\cite{DongL16} first considered tree-based models for code generation, which are generally used to ensure GC by deriving the syntactic structure \cite{RabinovichSK17}. To exploit the grammatical information of code, \cite{Yin17} adopted the encoder-decoder architecture to output an abstract syntax tree (AST) node sequence. On this basis, TRANX \cite{TranX} was proposed and became an effective and widely used tree-based model. Many works were based on and improved upon TRANX, such as ASED \cite{JiangSGMYS22}, Subtoken-TranX \cite{IndustryCodeGeneration}, and APT \cite{APT}. Moreover, \cite{SunZMXLZ19} and \cite{TreeGen} applied tree-based CNN and Transformer for code generation.

A comparative view of the tree-based and sequence-based approaches is shown in Table \ref{compare}. Although tree-based approaches can ensure GC, they have three major disadvantages: 1) AST node sequence is much longer (about 1.5-2 times) than token sequence \cite{APT}, leading to increased difficulty in generation. 2) The tree-based approach has to generate code with a complete tree structure, which makes the operation of obtaining arbitrary code snippets more difficult. 3) Assembling data for the tree-based approach is laborious because it requires the generation of AST based on syntactically complete code.

Compared to tree-based approaches, our proposed method not only ensures GC but also avoids the disadvantages of tree-based methods.


\begin{table}[ ht!]
	\caption{Comparisons of sequence-based methods and tree-based methods.}
	\centering
	\resizebox{0.47\textwidth}{!}{
		\begin{tabular}{ccc}
			\toprule
			& Sequence-based & Tree-based \\
			\midrule
                \multirow{2}{*}{Output length} & Same as token  & Much longer than token sequence \\
			& sequence of code &  of code (about 1.5-2 times)\\
			\midrule 
			Operation / NLP  & \multirow{2}{*}{Easy} & \multirow{2}{*}{Medium} \\
			technologies transfer & & \\
                 \midrule 
			\multirow{2}{*}{Data collection} & \multirow{2}{*}{Easy}         & Hard\\
			& & (Especially for large-scale data) \\
			\midrule 
			Grammatical & \multirow{2}{*}{No} & \multirow{2}{*}{Yes}\\
			correctness & &\\
			\bottomrule
	\end{tabular}}
    \label{compare}
\end{table}

\subsection{\mbox{Domain-specific Grammatical Sequence}-based Code Generation} \cite{PICARD} proposed PICARD that constrains auto-regressive decoders of language models through incremental parsing for a domain-specific language (DSL), i.e., Structured Query Language (SQL). PICARD used a parser for filtering in the beam search phase, meaning its candidate hypotheses might have terminal symbol mismatch errors. \cite{Synchromesh} proposed constrained semantic decoding (CSD) to address this problem for DSLs. However, both of them cannot extend to general-purpose languages (GPLs) like Python. For \cite{PICARD}, the cost of attempts could be overwhelming for large vocabulary, and GPLs usually have a much larger vocabulary than DSLs. For \cite{Synchromesh}, they use a intuitive method to deal with easy grammar of DSLs, but GPLs usually have more complicated grammar than DSLs, so as mentioned at the end of \cite{Synchromesh}, CSD cannot handle GPLs like Python.

There is a great need to ensure GC of sequence-based approaches for all PLs (including DSLs and GPLs), thus we introduce PDA to ensure GC for sequence-based code generation, which can be applied to all PLs.


\section{Methodology}
We first describe the core module of the proposed approach -- a PDA-based methodology consisting of a PDA module and the corresponding algorithm to obtain a valid set for ensuring GC during code generation.

\label{Methodology}

\begin{figure}[t!]
	\centering
	\includegraphics[width=0.5\textwidth]{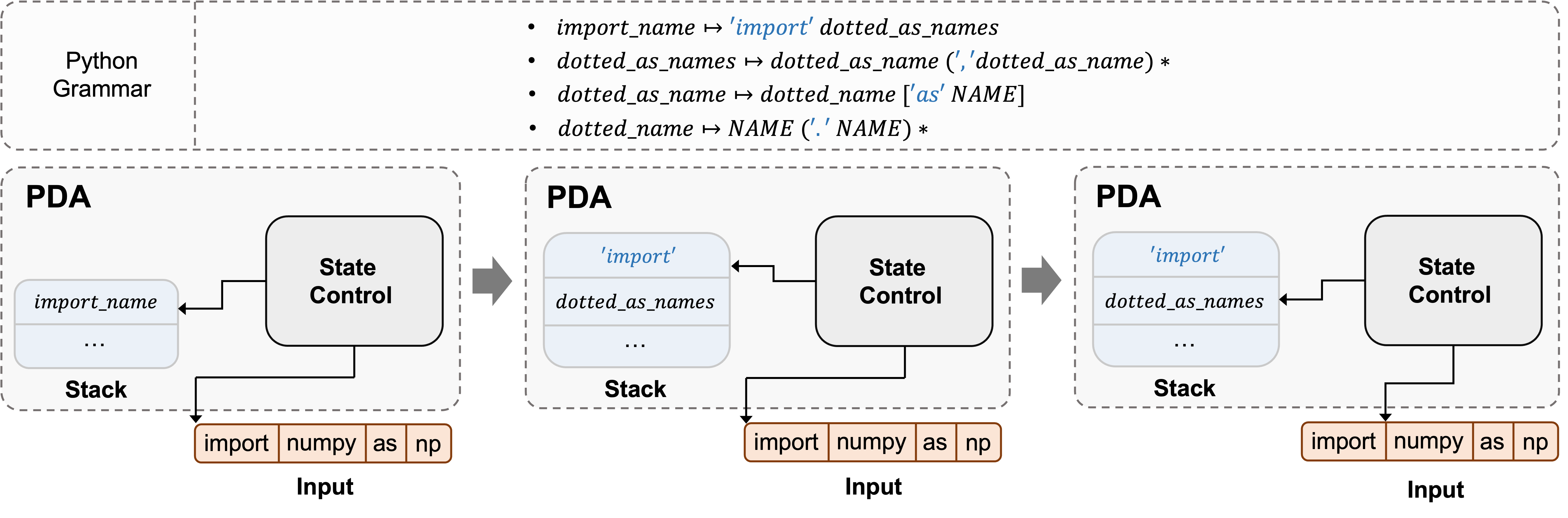}
	\caption{Schematic diagram of a Python grammar PDA parsing Python code.}
	\label{diagram}
\end{figure}

PLs belong to context-free languages, such as Python, Java, C++, etc., which can be accepted by a PDA \cite{chomsky1962context, schutzenberger1963context, evey1963application}. A PDA ${\displaystyle M}$ can be defined as:
\begin{equation}
	M=(S\,,\Sigma \,,\Gamma \,,s_{0}\,,g_{0}\,,A\,,\delta \,)
\end{equation}
where $S$ is a finite set of states, $\Sigma$ is a finite set of input symbols, $\Gamma$  is a finite set of stack symbols, $s_0 \in S$ is a start state, $g_0 \in \Gamma$ is a starting stack symbol, $A\subseteq S$, where $A$ is the set of accept states, and $\delta$ is a transition function, mapping $S\, \times \Gamma\, \times ( \Sigma\, \cup \left \{ \varepsilon\, \right \} )$  into the finite subsets of $S \times \Gamma ^{*}$, where $\displaystyle *$ is Kleene star. 

To construct a PDA module for a PL grammar, we first define $S$, $\Sigma$, $\Gamma$,  $s_0$, $g_0$, and $A$. In this paper, we adopt non-terminal symbols $\mathcal{N}'s$ of production rules in PL grammar as $S$, which can be changed with the PDA you build. $\Sigma$ is set to terminal symbols $\mathcal{T}'s$ of production rules and $\Gamma$ is the union of $\mathcal{T}'s$ and $\mathcal{N}'s$. In PL grammar, $s_0$ and $g_0$ are the starting non-terminal symbol, and $A$ is the set of ending terminal symbols. As an example, in Python grammar PDA, $s_0$ and $g_0$ can be chosen as `file\_input', and $A$ is the set of ENDMARKER token-type. Then, the definition of $\delta$ is:
\begin{itemize}
	\item For each $\mathcal{N}$, $\delta$(s, $\mathcal{N}$,$\epsilon$) = \{(s, $\beta$) | $\mathcal{N} \rightarrow \beta$ is a production rule in PL grammar\}.
	\item For each $\mathcal{T}$, $\delta$(s, $\mathcal{T}$, $\mathcal{T}$) = \{(s, $\epsilon$)\}.
\end{itemize}

The input symbols $\mathcal{I}'s$ in $\Sigma$ are finite, because grammar of PLs merges the same type of tokens. For instance, each $\mathcal{I}$ in Python belongs to one of 83 syntax-strings and 10 token-types. As shown in Fig. \ref{diagram}, `import' and `as' are syntax-strings, while `numpy' and `np' belong to NAME token-type.
Fig. \ref{diagram} illustrates a Python grammar PDA parsing `import numpy as np'. For a valid $\mathcal{I} =$ `import', PDA jumps to valid states and stacks based on $\delta$. 

As shown in Algorithm \ref{algorithm1}, given current state $s$ and current stack $g$, PDA ${\displaystyle M}$ is able to provide the set of valid $\mathcal{I}'s$ according to $\delta$. If $\varepsilon$ belongs to the set of valid $\mathcal{I}'s$, the state and stack transferred after entering $\varepsilon$ should be taken into account as well. Eventually, we merge all sets of $\mathcal{I}'s$ under each state and stack and remove $\varepsilon$ from them. 


\begin{algorithm}[htbp]
	\caption{Pseudocode for PDA module to obtain the valid set.}\label{algorithm1}
	\begin{algorithmic}[1]
		
		\REQUIRE{PDA ${\displaystyle M}$ and current state $s$ as well as stack $g$.}
		\ENSURE{The set of valid $\mathcal{I}'s$ and corresponding states and stacks.}
		
		\STATE Initial empty valid set $V$ and empty queue $Q$.
		\STATE Enqueue($Q$, $(s,g)$).
		\REPEAT
		\STATE $s, g \leftarrow$ Dequeue($Q$).
		\FOR{$(s, g, \mathcal{I}) \in \delta.keys()$} 
		\IF{$\mathcal{I} = \varepsilon$} 
		\STATE Enqueue($Q$, $\delta(s, g, \varepsilon)$).
		\ELSE 
		\STATE $V \leftarrow V \cup (\mathcal{I}, \delta(s, g, \mathcal{I}))$.
		\ENDIF 
		\ENDFOR 
		\UNTIL{$Q$ is empty}
		\RETURN{$V$}
	\end{algorithmic}
\end{algorithm}

\section{CodePAD}

In this section, we introduce CodePAD, which is constrained by the PDA module constructed under methodology. This framework can be applied to any sequence-based model, no matter it is an encoder-decoder or a decoder only. In this paper, we employ Transformer-based encoder-decoder model \cite{Transformer} as the backbone. We mainly modify the decode side, which adds PDA module, state representation, state prediction task, and joint prediction to help models generate grammatical code.

\begin{figure}[t!]
	\centering
	\includegraphics[width=0.5\textwidth]{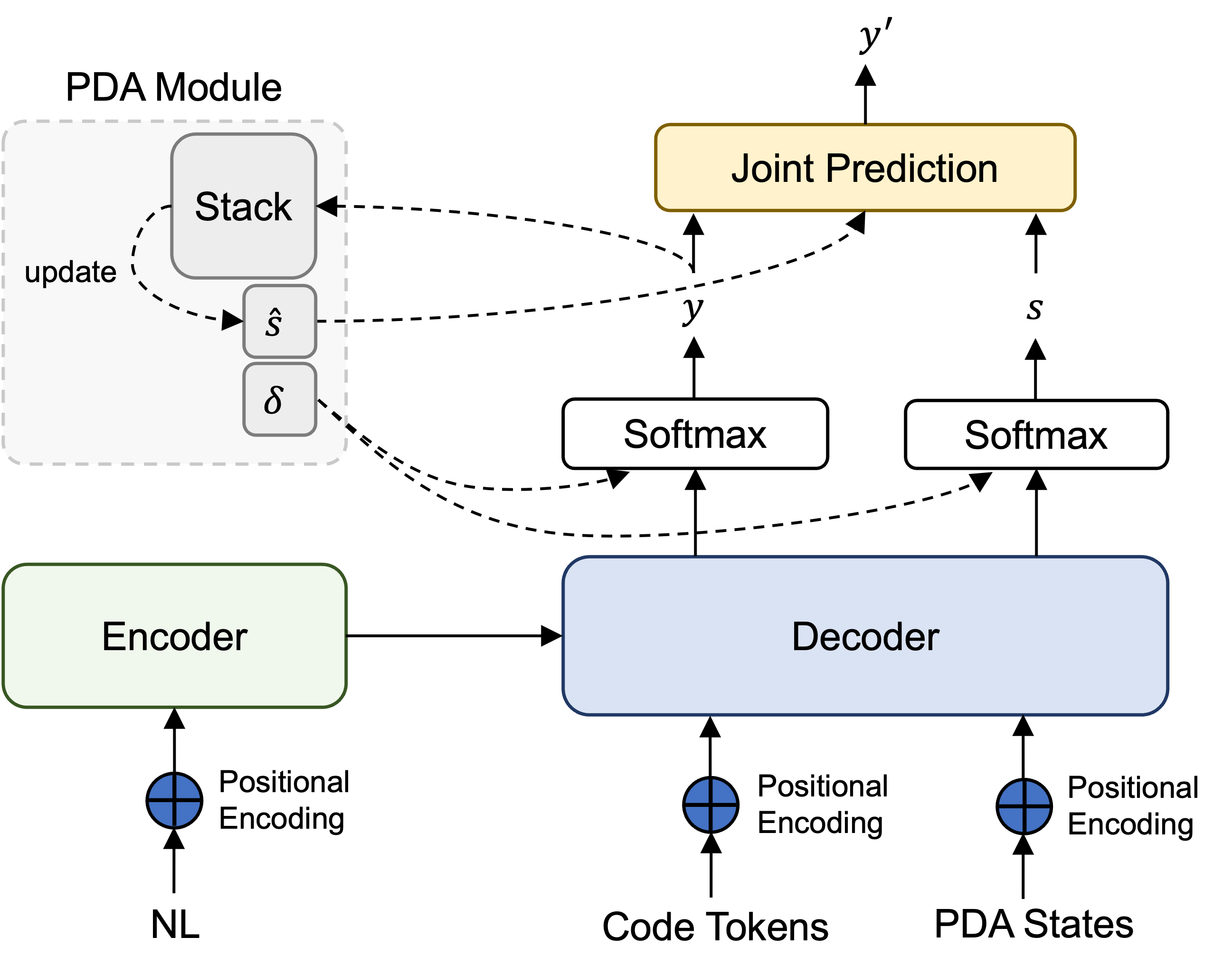}
	\caption{Diagram of CodePAD.}
	\label{architecture}
\end{figure}

Fig. \ref{architecture} shows the model architecture of CodePAD. At the encoder side of models, it maps an input sequence of NL utterances $\boldsymbol{x} = \left\{x_1,x_2,\cdots,x_n\right\}$ to a sequence
of continuous NL representations $\boldsymbol{z} = \left\{z_1,z_2,\cdots,z_n\right\}$. At the decoder side of models, given the NL representation sequence $\boldsymbol{z}$, it then generates a token sequence of code, one element at a time. At each step, models generate the next token in an auto-regressive manner \cite{Autoregressive}, which deploys previous generation results as additional input.

\subsection{State Representation}
The state is a type of valuable information to comprehend the status of the generated valid prefix in PDA module. In order to exploit states provided by PDA module, we set both tokens and corresponding states as the input of decoder. Specifically, given tokens $\boldsymbol{y}=\left\{y_1, y_2, \cdots, y_{t-1}\right\}$, we can obtain corresponding states $\boldsymbol{s}=\left\{s_1,s_2,\cdots,s_{t-1}\right\}$ from PDA module. Then, $\boldsymbol{y}$ and $\boldsymbol{s}$ are converted into $\boldsymbol{h}$ as follows:
\begin{align}
	& \boldsymbol{h}_y = \boldsymbol{e}_y + \operatorname{PE}(\boldsymbol{y})\\
	& \boldsymbol{h}_s = \boldsymbol{e}_s + \operatorname{PE}(\boldsymbol{s})\\
	& \boldsymbol{h}_t = \operatorname{Concat}(\boldsymbol{h}_y, \boldsymbol{h}_s)
\end{align}
where $\boldsymbol{e}_y$ and $\boldsymbol{e}_s$ are the embedding of $\boldsymbol{y}$ and $\boldsymbol{s}$ respectively, $\operatorname{Concat}(\boldsymbol{h}_y, \boldsymbol{h}_s)$ indicates the concatenation of $\boldsymbol{h}_y$ and $\boldsymbol{h}_s$, and $\operatorname{PE}$ indicates the positional encoding \cite{PositioalEncoding}.

\subsection{Token Prediction Task}
To generate code, $\boldsymbol{h}_t$ and the NL representation sequence $\boldsymbol{z}$ are fed into the decoder:
\begin{equation}
	\boldsymbol{a}_t = \operatorname{Decoder}(\boldsymbol{h}_t, \boldsymbol{z})
	\label{decoder}
\end{equation}

According to $\boldsymbol{a}_t$, we can compute the following probabilities:
\begin{align}
	& p(\operatorname{gen}  \mid \boldsymbol{a}_t) = \operatorname{softmax}(\textbf{W}_x\boldsymbol{a}_t), \\
	& p(\operatorname{copy} \mid \boldsymbol{a}_t) = 1 - p(\operatorname{gen}  \mid  \boldsymbol{a}_t),\\
	& p(v_c \mid \operatorname{gen},\boldsymbol{a}_t) = \operatorname{softmax}(\boldsymbol{e}_{v_c}^T\textbf{W}_v\boldsymbol{a}_t), \\
	& p(x_i \mid \operatorname{copy},\boldsymbol{a}_t,\boldsymbol{x})=\operatorname{softmax}(\boldsymbol{h}_{x_i}^T\textbf{W}_x\boldsymbol{a}_t).
\end{align}
where $\textbf{W}_g$, $\textbf{W}_v$, and $\textbf{W}_x$ are three different parameter matrices, $v_c$ indicates one of tokens in the vocabulary, and $\boldsymbol{h}_{x_i}$ is calculated by the pointer network \cite{PointerNetworks}. Finally, at t step, the probability of generated token $y_t$ can be defined as: 
\begin{align}
	& p(y_t  \mid  y_{<t}, s_{<t},\boldsymbol{x})  =  p(\operatorname{gen}  \mid \boldsymbol{a}_t)p(v_c \mid \operatorname{gen},\boldsymbol{a}_t) \nonumber \\
	& \qquad + p(\operatorname{copy} \mid \boldsymbol{a}_t)p(x_i \mid \operatorname{copy},\boldsymbol{a}_t,\boldsymbol{x}).
	\label{yt}
\end{align}

\begin{figure}[ht!]
	\centering
	\includegraphics[width=0.5\textwidth]{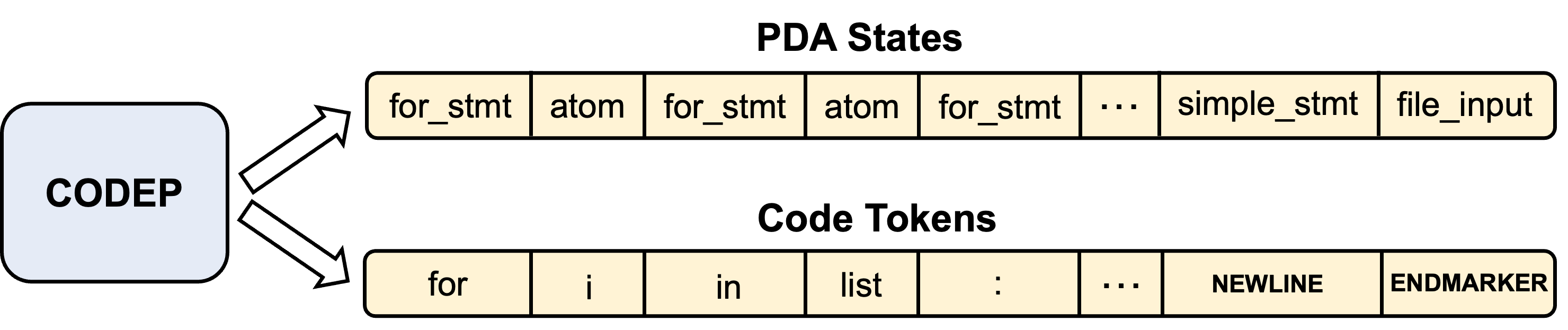}
	\caption{An example of state prediction task.}
	\label{example}
\end{figure}

\subsection{State Prediction Task}
To better understand the deduction process of PDA, we introduce an auxiliary task of PDA state prediction. Similar to token prediction, given $y_{<t}$, $s_{<t}$, and $\boldsymbol{x}$, we can calculate the probability of a state at $t$ step as follows:
\begin{equation}
	p(s_t \mid  y_{<t}, s_{<t},\boldsymbol{x}) = \operatorname{softmax}(\boldsymbol{e}_{v_s}^T\textbf{W}_s\boldsymbol{a}_t),
	\label{st}
\end{equation}
where $v_s$ indicates one of states in the vocabulary, $\textbf{W}_s$ is another parameter matrix, and $\boldsymbol{a}_t$ is calculated via \eqref{decoder}. Fig. \ref{example} shows an example of state prediction task. We can see that state prediction task also gives guidance for generating tokens. For example, `for', `in', and `:' are the fixed pairing of `for\_stmt' in Python.

\begin{table*}[htbp]
	\caption{Comparison of CodePAD and tree-based methods.}	\label{wp}
	\centering
        \resizebox{\textwidth}{!}{
	\begin{tabular}{lcccccc}
		\toprule
		\multirow{2}{*}{Model} & \multicolumn{3}{c}{CONALA}&\multicolumn{1}{c}{DJANGO} &\multicolumn{2}{c}{JUICE-10K}\\
            \cmidrule(r){2-4} \cmidrule(r){5-5} \cmidrule(r){6-7}
		& BLEU & CodeBLEU & EM & EM & BLEU & CodeBLEU\\
		\midrule
		\textbf{Tree-based methods w/o pre-training} & & & & & &\\
		TRANX \cite{Yin19Reranking} & $24.35 \pm 0.4$ & $26.80 \pm 0.6$ & $2.5 \pm 0.7$  & $77.3 \pm 0.4$ & $4.63 \pm 0.2$ & $13.09 \pm 0.4$ \\
		ML-TRANX \cite{Xie21MLTranX} & $24.42 \pm 0.8$ & $27.59 \pm 1.0$ & $2.2 \pm 0.4$  & $79.6 \pm 0.3$ & $4.75 \pm 0.4$ &  $13.43 \pm 0.5$\\
		TRANX-RL \cite{Jiang21TranXRL} & $25.47 \pm 0.7$ & $25.14 \pm 1.0$ & $2.6 \pm 0.4$ &  $77.9 \pm 0.5$ & $6.08 \pm 0.3$ &  $13.78 \pm 0.6$ \\
		APT \cite{APT} & $27.56 \pm 0.7$ & $28.54 \pm 0.7$ & $2.9 \pm 0.7$  & $79.3 \pm 0.5$ & $5.78 \pm 0.5$ & $14.19 \pm 0.5$\\
		Transformer-AST & $26.50 \pm 0.4$ & $27.31 \pm 0.4 $ & $1.8 \pm 0.2$ & $77.1 \pm 0.5$ & $4.80 \pm 0.3$&  $13.57 \pm 0.4$\\
		\midrule
		LSTM & $24.23 \pm 0.7$ & $25.76 \pm 0.8$ & $2.0 \pm 0.7$ & $70.5 \pm 0.4$ & $4.54 \pm 0.5$ & $13.13 \pm 0.7$\\ 
		CodePAD (LSTM-based) & \bm{$28.22 \pm 0.8$} & \bm{$30.12 \pm 0.9 \,(\uparrow 17\%)$} & \bm{$3.0 \pm 0.6$} & $79.6 \pm 0.6 \,(\uparrow 13\%)$& $6.32 \pm 0.6$ & $15.24 \pm 0.6 \,(\uparrow 16\%)$\\
		\hdashline
		Transformer & $24.21 \pm 0.8$ & $25.40 \pm 0.9$ & $2.0 \pm 0.6$ & $74.1 \pm 0.7$ & $4.67 \pm 0.4$& $14.08 \pm 0.5$ \\
		CodePAD (Transformer-based) &  $27.87 \pm 0.9$ & $29.49 \pm 0.9 \,(\uparrow 16\%)$ & $2.6 \pm 0.8$ & \bm{$79.7 \pm 0.5 \,(\uparrow 8\%)$ } & \bm{$7.26 \pm 0.5$}& \bm{$16.09 \pm 0.6 \,(\uparrow 15\%)$}\\
		\bottomrule
	\end{tabular}}
\end{table*}

\subsection{Training and Inference}
The purpose of the training stage is to minimize the sum of cross-entropy loss for two tasks, which is defined as:
\begin{align}
 	& \mathcal{L}^{\operatorname{ce}}(\boldsymbol{x}, \boldsymbol{y}, \boldsymbol{s}; \boldsymbol{\theta})  = \mathcal{L}^{\operatorname{ce}}_y(\boldsymbol{x}, \boldsymbol{y}, \boldsymbol{s}; \boldsymbol{\theta}) \\ \nonumber
	& \qquad \qquad  \qquad \quad + \alpha \cdot \mathcal{L}^{\operatorname{ce}}_s(\boldsymbol{x}, \boldsymbol{y}, \boldsymbol{s}; \boldsymbol{\theta}), \\
	&\mathcal{L}^{\operatorname{ce}}_y(\boldsymbol{x}, \boldsymbol{y}, \boldsymbol{s}; \boldsymbol{\theta})=-\sum_{t=1}^{T} \log p\left(y_{t} \mid y_{<t}, s_{<t}, \boldsymbol{x}; \boldsymbol{\theta}\right),  \\ 
	&\mathcal{L}^{\operatorname{ce}}_s(\boldsymbol{x}, \boldsymbol{y}, \boldsymbol{s}; \boldsymbol{\theta})=-\sum_{t=1}^{T} \log p\left(s_{t} \mid y_{<t}, s_{<t}, \boldsymbol{x}; \boldsymbol{\theta}\right), 
    \label{ce1}
\end{align}
where $\boldsymbol{\theta}$ is CodePAD's parameter, 
$\mathcal{L}^{\operatorname{ce}}_y$ and $\mathcal{L}^{\operatorname{ce}}_s$ are losses of token prediction task and state prediction task, respectively, and $\alpha$ is used to control the impact of them.

In the inference stage, we are able to calculate $p(y_t  \mid  y_{<t}, s_{<t},\boldsymbol{x})$ via \eqref{yt} with additional constraints of PDA module $y_t \in \{\mathcal{I} | (\mathcal{I}, s, g) \in V\}$, where $V$ is obtained according to Algorithm \ref{algorithm1}. Similarly, we can calculate $p(s_t  \mid  y_{<t}, s_{<t},\boldsymbol{x})$ via \eqref{st}. 
\subsubsection{Joint Prediction with State}
We can leverage both probabilities to jointly predict the result as follows:
\begin{align}
	& p(y'_t  \mid  y_{<t}, s_{<t},\boldsymbol{x})  = \frac{1}{1+\alpha}\cdot (p(y_t  \mid  y_{<t}, s_{<t},\boldsymbol{x}) \nonumber \\
	& \qquad + \frac{\alpha}{1+\alpha} \cdot p(\hat{s_t}  \mid  y_{<t}, s_{<t},\boldsymbol{x})),
     \label{y't}
\end{align}
where $\hat{s_t}$ indicates the actual state of PDA module, which is obtained from $\delta(s_{t-1}, g_{t-1}, y_t)$. Joint prediction wants to prevent CodePAD from generating tokens of inappropriate PDA states.


\section{Experiment Setup}
\label{Experiment Setup}

\begin{algorithm}[htbp!]
	\caption{Inference procedure of CodePAD.}\label{algorithm2}
	\begin{algorithmic}[1]
		
		\REQUIRE{The PDA ${\displaystyle M}$, hyperparameter $\alpha$, parameters of CodePAD $\boldsymbol{\theta}$, and NL utterances $\boldsymbol{x}$.}
		\ENSURE{The generated tokens $\boldsymbol{y'}$.}
		
		\STATE Initial $t \leftarrow 0$ and obtain $s_0$ and $g_0$.
		\REPEAT
		\STATE Obtain $V$ according to Algorithm \ref{algorithm1}.
		\STATE Calculate $p(y_t  \mid  y_{<t}, s_{<t},\boldsymbol{x})$ via \eqref{yt}, where \\ $y_t \in \{\mathcal{I} | (\mathcal{I}, s, g) \in V\}$. 
		\STATE Calculate $p(s_t \mid  y_{<t}, s_{<t},\boldsymbol{x})$ via \eqref{st}.
		\STATE Calculate $p(y'_t  \mid  y_{<t}, s_{<t},\boldsymbol{x})$ via \eqref{y't}.
		\STATE $y'_t \leftarrow \arg \max p(y'_t  \mid  y_{<t}, s_{<t},\boldsymbol{x})$.
		\STATE $y_t \leftarrow y'_t$.
		\STATE $s_{t+1}, g_{t+1} = \delta(s_t, g_t, y'_t)$ 
		\STATE $t \leftarrow t+1$
		\UNTIL{$s_t \in A$}
		\RETURN{$\boldsymbol{y'}$}
	\end{algorithmic}
\end{algorithm}

\subsection{PL and Datasets}
We build a PDA for the most popular PL Python (both Python2 and Python3) and conduct experiments on four public benchmark datasets, i.e., CONALA \cite{YinDCVN08}, DJANGO \cite{oda2015learning}, JUICE-10K \cite{Agashe19Juice}, and MBPP \cite{MBPP}.

\begin{table*}[tbp]
	\caption{Ablation study of CodePAD, where SR means state representation, SPT means state prediction task, and JP means joint prediction.}	\label{ab}
	\centering
	\scriptsize{
	\begin{tabular}{lcccccccc}
		\toprule
		\multirow{2}{*}{Model}& \multicolumn{3}{c}{CONALA} & \multicolumn{2}{c}{DJANGO} &\multicolumn{3}{c}{JUICE-10K}\\
            \cmidrule(r){2-4} \cmidrule(r){5-6} \cmidrule(r){7-9}
		& BLEU & CodeBLEU & GCP & EM & GCP & BLEU & CodeBLEU & GCP\\
		\midrule
		CodePAD (LSTM-based) & $28.22$ & $30.12$ & $100\%$ & $79.6$ & $100\%$ & $6.32$ & $15.24$ & $100\%$\\
		-JP & $27.81$ & $29.54$ & $100\%$ & $78.7$ & $100\%$ & 5.96 & 15.04 & $100\%$\\
		-SPT & $27.36$ & $28.95$ & $100\%$ & $77.3$ & $100\%$ & 5.38 & 14.26 & $100\%$\\
		-SR & $27.97$ & $29.98$ & $100\%$ & $77.6$ & $100\%$ & 5.54 & 14.47 & $100\%$\\
		-PDA-JP & $26.37$ & $27.49$ & $80.8\%$ & $74.1$ & $91.1\%$ & 5.09 & 13.87 & $54.7\%$ \\
		-PDA-SR-SPT-JP & $24.23$ & $25.76$ & $69.4\%$ & $70.5$ & $89.5\%$ & $4.54$ & $13.13$ & $43.9\%$ \\
		\midrule
		CodePAD (Transformer-based) & $27.87$ & $29.49$ & $100\%$ & $79.7$ & $100\%$ & $7.26$ & $16.09$ & $100\%$ \\
		-JP & $27.05$ & $28.85$ & $100\%$ & 78.9 & $100\%$ & $7.12$ & $15.82$ & $100\%$ \\
		-SPT & $26.72$ & $27.79$ & $100\%$ & 77.8 & $100\%$ & $6.59$ & $15.52$ & $100\%$ \\
		-SR & $27.13$ & $28.83$ & $100\%$ & 78.3 & $100\%$ & $6.97$ & $15.68$ & $100\%$ \\
		-PDA-JP & $26.32$ & $27.46$ & $81.4\%$ & 76.7 & $92.3\%$ & $5.45$ & $15.20$ & $57.6\%$\\
		-PDA-SR-SPT-JP & $24.21$ & $25.40$ & $70.6\%$  & $74.1$ & $90.1\%$ & $4.67$ & $14.79$ & $45.3\%$\\
		\bottomrule
	\end{tabular}}
\end{table*}

\subsection{Baselines}
\label{baselines}
We select two typical \textbf{sequence-based models}, i.e., LSTM \cite{LSTM} and Transformer \cite{Transformer}, as our base models. 
For \textbf{non-pre-trained tree-based methods}, we compare CodePAD with TRANX \cite{Yin19Reranking}, ML-TRANX \citep{Xie21MLTranX}, TRANX-RL \citep{Jiang21TranXRL}, APT \cite{APT}, and Transformer-AST \footnote{Transformer-AST indicates a tree-based method based on Transformer instead of LSTM adopted in other tree-based methods mentioned above.}, which ensure GC of generated codes relying on AST. 
For \textbf{pre-trained sequence-based methods}, we choose encoder-decoder models, including BART \cite{BART} and CodeT5 \cite{CodeT5}, and decoder-only models, including CodeGen \cite{Codegen} and InCoder \cite{Incoder}.

\subsection{Evaluation Metrics}
We use three widely-used evaluation metrics in code generation: exact matching accuracy (EM), corpus-level BLEU-4 (BLEU), and CodeBLEU \cite{CodeBLEU}, which consider important grammar and semantic features of codes. We also use GC and GC percentage (GCP). GCP indicates the percentage of grammatically generated codes in all generated codes. 

\subsection{Implementation Details}
For approaches mentioned in Section \ref{baselines}, we follow the settings in their paper. To mitigate the instability of model training, we exhibit the average performance of models running five times. 
Details of datasets, hyperparameters, baselines, and research questions refer to Section \ref{details} in appendices.

\section{Experimental Results}
\label{Experimental Results}
We conduct experiments to answer the following research questions (RQs):

\subsection{RQ1: CodePAD vs Tree-based Methods}
From Tables \ref{wp}, we can observe that CodePAD substantially outperforms SOTA tree-based methods without pre-training on three public benchmark datasets in terms of BLEU, CodeBLEU, and EM. It indicates that although both tree-based methods and CodePAD generate grammatical codes, CodePAD takes advantage of the shorter generation sequence length of code tokens than AST nodes. We also find that Transformer-based CodePAD achieves competitive performance against LSTM-based CodePAD on CONALA. However, on DJANGO and JUICE-10K, Transformer-based CodePAD exhibits its superiority due to more extensive training data and longer generated token sequence. Therefore, for large dataset and long code generation, we recommend employing Transformer as the base model of CodePAD. Since the decoder of most tree-based methods without pre-training is based on LSTM, we implement Transformer-AST to verify that the effect of CodePAD is derived from our proposed method rather than the base model. The experimental results show that the base model is not the main factor affecting performance.

\begin{table*}[ht!]
	\caption{Examples of code generation for each pre-trained model in zero-shot setting, where NL is `Write a function to find the similar elements from the given two tuple lists' and GOLD denotes the reference code. }	\label{e}
	\centering
	\resizebox{\textwidth}{!}{
	\begin{tabular}{lll}
		\toprule
		Model & Code\\
		\midrule
		BART-125M & Write a function to find the similar elements from the given two tuple lists. \\
		\hdashline
		CodeT5-220M &  def f\_elements( )\\
		\hdashline
		CodeT5-770M & to  find  the  similar  elements  ." ," ." ," a  function  to  find  the  similar  elements  from  the  given  two  tuple  lists . \\
		\hdashline
		CodeGen-350M & def find\_similar ( t1 , t2 ): $\backslash$n """ $\backslash$n Finds the similar elements from the given two tuples. $\backslash$n $\backslash$n\\
		\hdashline
		InCoder-1B & def similar\_elements ( t1, t2 ): $\backslash$n $\backslash$n $\backslash$n $\backslash$n $\backslash$n $\backslash$n $\backslash$n $\backslash$n $\backslash$n $\backslash$n $\backslash$n $\backslash$n \\
		\midrule
		CodeGen-350M +PDA & def find ( a , b ): $\backslash$n for i in  range ( len ( a ) ): $\backslash$n if a [ i ] == b [ i ]: $\backslash$n return a[i]\\
		\hdashline
		GOLD & def similar\_elements ( test\_tup1, test\_tup2 ) : $\backslash$n res = tuple ( set ( test\_tup1 ) \& set ( test\_tup2 ) ) $\backslash$n return ( res )\\
		\bottomrule
	\end{tabular}}
\end{table*}

\subsection{RQ2: Ablation Study}
In Table \ref{ab}, we demonstrate the performance of LSTM-based and Transformer-based CodePAD with the reduction of some parts of them. The results indicate that each part of CodePAD contributes, and PDA module is the most critical part of CodePAD. When PDA module is removed, GCP drops sharply, which is inversely correlated with the average code length of datasets, and other evaluation metrics also have degradation. JP becomes unavailable due to the dependence of JP on the mapping from the token to the corresponding state provided by PDA module. In addition, it can be seen that only reducing JP or SR leads to a relatively small decrease in performance, because they have some overlap with other parts in helping the model. `-PDA-SR-SPT-JP' means that only the base model is used, which has a significant performance degradation compared to CodePAD (both LSTM-based and Transformer-based). As shown in Tables \ref{wp}, it relatively improves 17\% CodeBLEU on CONALA, 8\% EM on DJANGO, and 15\% CodeBLEU on JUICE-10K compared to base models.

\subsection{RQ3: Improvements of Sequence-based Methods with PDA Module}

It is apparent from Table \ref{p} that pre-trained models fail to work on MBPP in zero-shot setting, because they are working in unfamiliar territory. With the help of PDA module, pre-trained sequence-based models achieve significant improvements. The reason may be that pre-training models have the ability to answer questions, but do not know how to organize them into well-formed codes. Under the constraints of PDA module, pre-training models are guided to exert their capabilities by generating grammatical codes. Especially, (BLEU, CodeBLEU) of CodeGen-350M without prompt \footnote{We explore the performance of decoder-only models under conditions that no prompt is given or only "def" is given as prompt, because traditional prompt `def + function signature' contains much valid information, including function name, argument type and name, and even return value type.} improvement from (1.55, 3.21) to (14.44, 21.54) on MBPP in zero-shot setting. 

\subsection{RQ4: Case Study}
In Table \ref{e}, we demonstrate examples of code generation for each pre-trained model in zero-shot setting. We can see that they cannot generate grammatical code or even complete code without prompt in zero-shot setting.
For example, CodeGen-350M generates the function signature and some comments capturing semantic information, which shows the ability to generate complete code if under proper guidelines. Under the constraints of PDA module, CodeGen-350M generates grammatical code. Although the code generated by CodeGen-350M + PDA still falls short of GOLD, we still think it is an impressive result considering that it is under the zero-shot setting.

\begin{table}[ht!]
\caption{Zero-shot results of each pre-trained sequence-based method with PDA module.}

\label{p}
\centering
\resizebox{0.47\textwidth}{!}{
\begin{tabular}{lccc}
	\toprule
	\multirow{2}{*}{Model}&\multicolumn{3}{c}{MBPP (Zero-shot)}\\
        \cmidrule(r){2-4}
	& BLEU & CodeBLEU & GC \\
	\midrule
	\textbf{Encoder-decoder methods w/ pre-training}\\
	BART-125M & 0.02 & 3.74 & False\\
	+PDA & 4.83 & 9.17 & True\\
	\hdashline
	CodeT5-220M & 0.01 & 0.57 & False \\
	+PDA & 4.75 & $9.79$ & True\\
	\hdashline
	CodeT5-770M & 0.02 & 3.98 & False \\
	+PDA & $6.19$ & 9.31 & True\\
	\midrule
	\textbf{Decoder-only methods w/ pre-training}\\
	CodeGen-350M & 1.55 & 3.21 & False\\
	+prompt & 0.33 &10.13 & False \\
	+PDA & \textbf{14.44} & \textbf{21.54} & True\\ 
	\hdashline
	InCoder-1B & 0.12 & 7.41 & False\\
	+prompt & 0.82 & 7.22 & False\\
	+PDA & 9.84 & 17.38 & True \\
	\bottomrule
\end{tabular}}
\end{table}

\section{Limitation}
There are three major limitations of our work: 
\begin{itemize}
    \item First, a specific PDA should be built for each PL, and we only built PDA for Python (including both Python2 and Python3). However, PDA can recognize context-free language to which PL belongs, due to time and memory limitations of running PL \cite{salomaa2001half}. In future work, we will build PDAs for more PLs.
    \item Second, our work ensures GC, but neither our method nor other methods can prevent compilation errors. With our proposed training framework, our method outperforms other methods on compilation rates, e.g., CODEP has 99.2\% compilation rates after training on CONALA.
    \item Third, using PDA will increase CPU operation and memory overhead, but it is still acceptable.
\end{itemize}

\section{Conclusion and Discussion}
\label{Conclusion and Discussion}
In this paper, we have proposed a sequence-based framework, namely CodePAD, that first integrates the deduction of PDA into deep learning for code generation. CodePAD adds PDA module, state representation, state prediction task, and joint prediction for the decoder of CodePAD. PDA module guarantees the GC of generated codes, while the others make use of valid information provided by PDA module (i.e., PDA state) to generate codes better. As a result, CodePAD dramatically enhances the performance of base models and outperforms SOTA tree-based methods without pre-training on three benchmark datasets. The ablation study demonstrates each component of CodePAD contributes. With the help of PDA module, pre-trained models also achieve significant improvements. 

PDA module shows excellent potential in code generation tasks in zero-shot setting, which can help pre-trained models apply to niche PLs that have little data. Furthermore, PDA can accommodate context-free language to satisfy grammar constraints, not just PL. We hope this work sheds light on future work in this direction. 



\newpage

\onecolumn
\appendix
\section{Syntax-strings and Token-types in Python}
In Table \ref{sstt}, we show each of 83 syntax-strings and 10 token-types in Python.

\begin{table}[ht!]
	\caption{Terminal symbols of Python grammar.} \label{sstt}
	\centering
	\begin{tabular}{ll}
		\toprule
		 & \multicolumn{1}{c}{Python}\\
		\midrule
		\multirow{6}{*}{Syntax-string} & `for', `in', `try', `finally', `with', `except', `lambda', `or', `and', `not', `del', `pass',\\ & `break', `continue', `return', `raise', `from', `import', `as', `nonlocal', `global',\\ & `assert', `if', `else', `elif', `while', `async', `def', `\char`@', `\char`(', `\char`)', `\char`-\char`>', `\char`:', `\char`*\char`*', `\char`*', `\char`,',\\ & `\char`=', `\char`;', `\char`^\char`=', `\char`\%\char`=', `\char`/\char`/\char`=', `\char`@\char`=', `<<=', `\char`*\char`*\char`=', `\char`\&\char`=', `\char`*\char`=', `\char`|\char`=', `\char`>\char`>\char`=', `\char`-\char`=', `\char`+\char`=', `\char`\/\char`=',\\ & `\char`.',  `...', `<=', `>=', `is', `==', `<', `>', `<>', `!=', `\char`|', `\char`^', `\&', `<<', `>>', `+', `-', `\%',\\ & `/', `//', `\char`~', `\char`!', `await', `False', `[', `\{', `True', `None', `]', `\}', `class', `yield'\\
		\midrule
		\multirow{2}{*}{Token-type} & `NAME', `STRING', `NUMBER', `INDENT', `DEDENT', `FSTRING\_START', \\& `FSTRING\_END', 
		`FSTRING\_STRING', `NEWLINE', `ENDMARKER' \\
		\bottomrule
	\end{tabular}
\end{table}

\section{Deterministic Pushdown Automaton}
Due to computer memory constraints, most of PLs are deterministic context-free languages, which can be accepted by a deterministic pushdown automaton (DPDA) \cite{salomaa2001half}. A PDA ${\displaystyle M}$ is deterministic only if both the following two conditions are satisfied:
\begin{itemize}
	\item $\forall s \in S, g \in \Gamma, \mathcal{I} \in \Sigma \cup \left \{ \varepsilon \right \}$, $|\delta(s,g,\mathcal{I})| = 1$ where $|\delta(s,g,\mathcal{I})|$ indicates the element number of the set $\delta(s,g,\mathcal{I})$.
	\item $\forall s \in S, g \in \Gamma, \mathcal{I} \in \Sigma$, if $\delta(s, g, \varepsilon) \not= \emptyset$, then $\delta\left(s,g,\mathcal{I} \right) = \emptyset$.
\end{itemize}

For the same deterministic context-free grammar, the general PDA is more accessible to construct than the DPDA. In addition, deterministic context-free languages are a proper subset of context-free languages. Therefore, in this paper, our proposed method is based on the general PDA, which is also applicable to the DPDA.

\section{Elementary Knowledge of Model Architecture}
The positional encoding \cite{PositioalEncoding} is defined as:
\begin{align*}
	\operatorname{PE}_{(pos, 2j)} &=\sin \left(pos / 10000^{2 j / d}\right) \\
	\operatorname{PE}_{(pos, 2j+1)} &=\cos \left(pos / 10000^{2 j / d}\right)
\end{align*}
where $pos$ is the position, $j$ is the dimension, and $d$ is the number of dimensions (i.e., embedding size). 

The output of the multi-headed self-attention in the decoder layer is computed via: 
\begin{align}
	&\bm{Q_i} =\bm{H} \bm{W_i^{Q}}, \bm{K}=\bm{H} \bm{W_i^{K}}, \bm{V}=\bm{H} \bm{W_i^{V}},\\
	&\bm{head_i} = \operatorname{softmax}\left(\frac{\bm{Q_i} \bm{K_i}^{\top}}{\sqrt{d_{k}}}\right) \bm{V_i}, \\
	&\bm{head} = \operatorname{Concat}(head_1, head_2, ..., head_h)\bm{W^{O}}
\end{align}
Where $\bm{W_i^{Q}} \in {\mathbb{R}}^{d_x \times d_k}, \bm{W^{K}} \in {\mathbb{R}}^{d_x \times d_k}, \bm{W^{V}} \in {\mathbb{R}}^{d_x \times d_h}$ are learnable parameter matrices. After that, a feed-forward layer and layer normalization are followed.

\section{Details of Datasets, Hyperparameters, Baselines, and Research Questions}
\label{details}
\subsection{Datasets}
\begin{itemize}
\item \textbf{CONALA} \cite{YinDCVN08} contains 2879 real-world data of manually annotated NL questions and their Python3 solutions on STACK OVERFLOW.
\item \textbf{DJANGO} \cite{oda2015learning} contains 18805 NL-annotated python2 code extracted from the Django Web framework.
\item \textbf{JUICE-10K} contains 10K training samples randomly selected from the training set of JUICE \cite{Agashe19Juice}, and the validation and test sets of JUICE-10K are consistent with those of JUICE. Due to the high demand for training resources, we use a subset instead of the full JUICE.
\item \textbf{MBPP} \cite{MBPP} contains 974 Python programming problems, and each sample consists of an NL, a code, and 3 test cases.
\end{itemize}  

\begin{table}[ht!]
	\caption{Statistics of datasets.}	\label{statistics}
	\centering
        \small{
	\begin{tabular}{llllll}
		\toprule
		 \multirow{2}{*}{Dataset} &\multicolumn{3}{c}{Examples Num}&\multicolumn{2}{c}{Avg Length} \\
            \cmidrule(r){2-4} \cmidrule(r){5-6}
		   & Train & Dev & Test & NL & Code \\
		\midrule
		CONALA & 2175 & 200 & 500 & 10.2 & 15.1\\
		DJANGO & 16000 & 1000 & 1805 & 14.1 & 10.6\\
		JUICE-10K & 10000 & 1831 & 2115 & 40.4 & 43.4 \\
		MBPP & - & - & 974 & 16.0 & 33.0\\
		\bottomrule
	\end{tabular}}
\end{table} 

\subsection{Hyperparameters}
We train our model with Adam \cite{adam} optimizer on a single GPU of Tesla A100-PCIe-40G. We set sizes of the word embedding, code embedding, state embedding, and hidden state as 256, 256, 256, and 512, respectively. The learning rate is set to $0.0001$ for Transformer-based CodePAD and $0.001$ for LSTM-based CodePAD. For additional hyperparameter $\alpha$, we pick $\alpha \in [0, 1]$ on validation sets. The beam size is set to 2 for MBPP and 15 for the other datasets. 

\subsection{Baselines}
\subsubsection{Non-pre-trained Tree-based Methods} 
\begin{itemize}
\item \textbf{TRANX} \cite{Yin19Reranking} uses a tree-based model to generate the AST as the intermediate representation of code.
\item \textbf{ML-TRANX} \citep{Xie21MLTranX} adopts a mutual learning framework to train models for different traversals-based decodings jointly.
\item \textbf{TRANX-RL} \citep{Jiang21TranXRL} uses a context-based branch selector to determine optimal branch expansion orders for multi-branch nodes dynamically. 
\item \textbf{APT} \cite{APT} uses antecedent prioritized loss to help tree-based models attach importance to antecedent predictions by exploiting the position information of the generated AST nodes.
\end{itemize}
\subsubsection{Pre-trained Sequence-based Methods} 
\begin{itemize}
\item \textbf{BART} \cite{BART} is a pre-training approach for text generation tasks that learns to map corrupted documents to the original.
\item \textbf{CodeT5} \cite{CodeT5} is a unified pre-trained encoder-decoder Transformer model that makes better use of the code semantics conveyed from developer-assigned identifiers.
\item \textbf{CodeGen} \cite{Codegen} is a unified generation model that allows left-to-right code generation and code infilling/editing by the causal mask language modeling training objective.
\item \textbf{InCoder} \cite{Incoder} is a series of large-scale language models trained on NL and programming data for conversation-based program synthesis. 
\end{itemize}
\subsection{Research Questions}
We conduct experiments to answer the following research questions (RQs):
\begin{itemize}
	\item RQ1: How well does the proposed CodePAD perform compared to the SOTA tree-based methods, which guarantees the syntactical correctness of generated codes relying on AST, without pre-training?
	\item RQ2: How does each part of our proposed method contribute to CodePAD?
	\item RQ3: How does the proposed PDA module improve the effectiveness of the SOTA pre-trained sequence-based methods?
	\item RQ4: How does CodePAD generate grammatical codes with the help of PDA module?
\end{itemize}

\begin{figure}[ht!]
	\centering
	\includegraphics[width=0.95\textwidth]{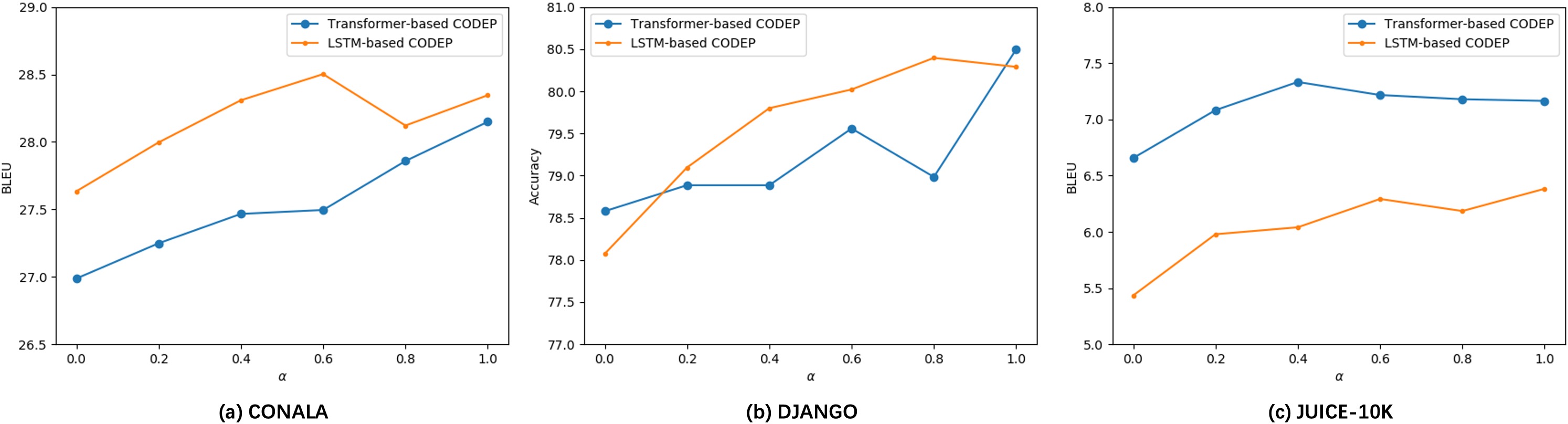}
	\caption{Effects of $\alpha$ on the validation sets.}
	\label{alpha}
\end{figure}

\section{Effect of $\alpha$}
The coefficient $\alpha$ is an important hyperparameter that controls the relative impacts of the state prediction task in the training stage and that of the joint prediction with the state probability in the inference stage. Therefore, we investigate the effect of $\alpha$ on our proposed framework in Fig. \ref{alpha}, which varies $\alpha$ from 0 to 1 with an increment of 0.2 on the validation set of CONALA, DJANGO, and JUICE-10K datasets. The experimental results show that as $\alpha$ increases, the performance of CodePAD increases when $\alpha \leq 0.4$, and its tendency is related to the base model and the dataset when $\alpha > 0.4$. What stands out in Fig. \ref{alpha} is that CodePAD performs relatively well on each of the above datasets when $\alpha$ is set to 1.0, which indicates the effectiveness of the state prediction task and joint prediction. For experiments in this paper, we set $\alpha$ as the optimal $\alpha$ in Fig. \ref{alpha}. In particular, the $\alpha's$ of the LSTM-based CodePAD and the transformer-based CodePAD are set to (0.6, 1.0), (0.8, 1.0), and  (1.0, 0.4) on CONALA, DJANGO, and JUICE-10K dataset, respectively.

\begin{table}[ht!]
	\caption{Three types of compilation error generated by CodePAD on CONALA dataset.}	\label{compilation error}
	\centering
        \small{
	\begin{tabular}{ll}
		\toprule
		Type & Example\\
		\midrule
		\multirow{2}{*}{I} & \multirow{2}{*}{1 = [ ( i , 2 ) for i in range(1) ]}\\
		\\
		\hdashline
		\multirow{2}{*}{II} & \multirow{2}{*}{plt.plot(x, var1, color=`str0', color=`str1')}\\
		\\
		\hdashline
		\multirow{2}{*}{III} & html = response.read(str0, `str0') \\
		& slot\_map=\{str0: http://www.example.com/\}\\
		\bottomrule
	\end{tabular}}
\end{table}

\section{Compilation Error Analysis}
Program execution needs to be compiled correctly in addition to being grammatically correct, and GC is the foundation of compilation. The compilation rate of CodePAD on CONALA dataset is $99.2\%$, and Table \ref{compilation error} demonstrates examples of three types of compilation error. The first type of compilation error is because the number type cannot be assigned, the second type of compilation error is the error of the duplicate parameter, and the third type of compilation error is the mapping error. We found that the above types of errors also occur in tree-based methods. For the first two types of errors, we are able to modify the transfer function, while the last type of error can be avoided using engineering methods.

\newpage
\twocolumn
\bibliography{acl}

\begin{thebibliography}{41}
\expandafter\ifx\csname natexlab\endcsname\relax\def\natexlab#1{#1}\fi

\bibitem[{Agashe et~al.(2019)Agashe, Iyer, and Zettlemoyer}]{Agashe19Juice}
Rajas Agashe, Srinivasan Iyer, and Luke Zettlemoyer. 2019.
\newblock Juice: {A} large scale distantly supervised dataset for open domain
  context-based code generation.
\newblock In \emph{{EMNLP/IJCNLP} {(1)}}, pages 5435--5445.

\bibitem[{Ahmad et~al.(2021)Ahmad, Chakraborty, Ray, and
  Chang}]{AhmadCRC21PLBART}
Wasi~Uddin Ahmad, Saikat Chakraborty, Baishakhi Ray, and Kai{-}Wei Chang. 2021.
\newblock Unified pre-training for program understanding and generation.
\newblock In \emph{{NAACL-HLT}}, pages 2655--2668. Association for
  Computational Linguistics.

\bibitem[{Austin et~al.(2021)Austin, Odena, Nye, Bosma, Michalewski, Dohan,
  Jiang, Cai, Terry, Le, and Sutton}]{MBPP}
Jacob Austin, Augustus Odena, Maxwell~I. Nye, Maarten Bosma, Henryk
  Michalewski, David Dohan, Ellen Jiang, Carrie~J. Cai, Michael Terry, Quoc~V.
  Le, and Charles Sutton. 2021.
\newblock Program synthesis with large language models.
\newblock \emph{CoRR}, abs/2108.07732.

\bibitem[{Cao et~al.(2019)Cao, Zhu, Liu, Li, and Yu}]{Cao19}
Ruisheng Cao, Su~Zhu, Chen Liu, Jieyu Li, and Kai Yu. 2019.
\newblock Semantic parsing with dual learning.
\newblock In \emph{{ACL} {(1)}}, pages 51--64.

\bibitem[{Chomsky(1962)}]{chomsky1962context}
Noam Chomsky. 1962.
\newblock Context-free grammars and pushdown storage.
\newblock \emph{MIT Res. Lab. Electron. Quart. Prog. Report.}, 65:187--194.

\bibitem[{Dong and Lapata(2016)}]{DongL16}
Li~Dong and Mirella Lapata. 2016.
\newblock Language to logical form with neural attention.
\newblock In \emph{{ACL} {(1)}}.

\bibitem[{Dong et~al.(2022)Dong, Li, and Jin}]{APT}
Yihong Dong, Ge~Li, and Zhi Jin. 2022.
\newblock Antecedent predictions are dominant for tree-based code generation.
\newblock \emph{CoRR}, abs/2208.09998.

\bibitem[{Evey(1963)}]{evey1963application}
R~James Evey. 1963.
\newblock Application of pushdown-store machines.
\newblock In \emph{Proceedings of the November 12-14, 1963, fall joint computer
  conference}, pages 215--227.

\bibitem[{Fried et~al.(2022)Fried, Aghajanyan, Lin, Wang, Wallace, Shi, Zhong,
  Yih, Zettlemoyer, and Lewis}]{Incoder}
Daniel Fried, Armen Aghajanyan, Jessy Lin, Sida Wang, Eric Wallace, Freda Shi,
  Ruiqi Zhong, Wen{-}tau Yih, Luke Zettlemoyer, and Mike Lewis. 2022.
\newblock Incoder: {A} generative model for code infilling and synthesis.
\newblock \emph{CoRR}, abs/2204.05999.

\bibitem[{Gehring et~al.(2017)Gehring, Auli, Grangier, Yarats, and
  Dauphin}]{PositioalEncoding}
Jonas Gehring, Michael Auli, David Grangier, Denis Yarats, and Yann~N. Dauphin.
  2017.
\newblock Convolutional sequence to sequence learning.
\newblock In \emph{{ICML}}, volume~70 of \emph{Proceedings of Machine Learning
  Research}, pages 1243--1252. {PMLR}.

\bibitem[{Graves(2013)}]{Autoregressive}
Alex Graves. 2013.
\newblock Generating sequences with recurrent neural networks.
\newblock \emph{CoRR}, abs/1308.0850.

\bibitem[{Guo et~al.(2022)Guo, Lu, Duan, Wang, Zhou, and Yin}]{UniXcoder}
Daya Guo, Shuai Lu, Nan Duan, Yanlin Wang, Ming Zhou, and Jian Yin. 2022.
\newblock Unixcoder: Unified cross-modal pre-training for code representation.
\newblock In \emph{{ACL} {(1)}}, pages 7212--7225. Association for
  Computational Linguistics.

\bibitem[{Hochreiter and Schmidhuber(1997)}]{LSTM}
Sepp Hochreiter and J{\"{u}}rgen Schmidhuber. 1997.
\newblock Long short-term memory.
\newblock \emph{Neural Comput.}, 9(8):1735--1780.

\bibitem[{Jia and Liang(2016)}]{JiaL16}
Robin Jia and Percy Liang. 2016.
\newblock Data recombination for neural semantic parsing.
\newblock In \emph{{ACL} {(1)}}.

\bibitem[{Jiang et~al.(2022)Jiang, Song, Ge, Meng, Yao, and Su}]{JiangSGMYS22}
Hui Jiang, Linfeng Song, Yubin Ge, Fandong Meng, Junfeng Yao, and Jinsong Su.
  2022.
\newblock An {AST} structure enhanced decoder for code generation.
\newblock \emph{{IEEE} {ACM} Trans. Audio Speech Lang. Process.}, 30:468--476.

\bibitem[{Jiang et~al.(2021)Jiang, Zhou, Meng, Zhang, Zhou, Huang, Wu, and
  Su}]{Jiang21TranXRL}
Hui Jiang, Chulun Zhou, Fandong Meng, Biao Zhang, Jie Zhou, Degen Huang,
  Qingqiang Wu, and Jinsong Su. 2021.
\newblock Exploring dynamic selection of branch expansion orders for code
  generation.
\newblock In \emph{{ACL/IJCNLP}}, pages 5076--5085.

\bibitem[{Kingma and Ba(2015)}]{adam}
Diederik~P. Kingma and Jimmy Ba. 2015.
\newblock Adam: {A} method for stochastic optimization.
\newblock In \emph{{ICLR}}.

\bibitem[{Lewis et~al.(2020)Lewis, Liu, Goyal, Ghazvininejad, Mohamed, Levy,
  Stoyanov, and Zettlemoyer}]{BART}
Mike Lewis, Yinhan Liu, Naman Goyal, Marjan Ghazvininejad, Abdelrahman Mohamed,
  Omer Levy, Veselin Stoyanov, and Luke Zettlemoyer. 2020.
\newblock {BART:} denoising sequence-to-sequence pre-training for natural
  language generation, translation, and comprehension.
\newblock In \emph{{ACL}}, pages 7871--7880. Association for Computational
  Linguistics.

\bibitem[{Ling et~al.(2016)Ling, Blunsom, Grefenstette, Hermann, Kocisk{\'{y}},
  Wang, and Senior}]{Ling16}
Wang Ling, Phil Blunsom, Edward Grefenstette, Karl~Moritz Hermann, Tom{\'{a}}s
  Kocisk{\'{y}}, Fumin Wang, and Andrew~W. Senior. 2016.
\newblock Latent predictor networks for code generation.
\newblock In \emph{{ACL} {(1)}}.

\bibitem[{Nijkamp et~al.(2022)Nijkamp, Pang, Hayashi, Tu, Wang, Zhou, Savarese,
  and Xiong}]{Codegen}
Erik Nijkamp, Bo~Pang, Hiroaki Hayashi, Lifu Tu, Huan Wang, Yingbo Zhou, Silvio
  Savarese, and Caiming Xiong. 2022.
\newblock A conversational paradigm for program synthesis.
\newblock \emph{CoRR}, abs/2203.13474.

\bibitem[{Oda et~al.(2015)Oda, Fudaba, Neubig, Hata, Sakti, Toda, and
  Nakamura}]{oda2015learning}
Yusuke Oda, Hiroyuki Fudaba, Graham Neubig, Hideaki Hata, Sakriani Sakti,
  Tomoki Toda, and Satoshi Nakamura. 2015.
\newblock Learning to generate pseudo-code from source code using statistical
  machine translation.
\newblock In \emph{2015 30th IEEE/ACM International Conference on Automated
  Software Engineering (ASE)}, pages 574--584. IEEE.

\bibitem[{Poesia et~al.(2022)Poesia, Polozov, Le, Tiwari, Soares, Meek, and
  Gulwani}]{Synchromesh}
Gabriel Poesia, Oleksandr Polozov, Vu~Le, Ashish Tiwari, Gustavo Soares,
  Christopher Meek, and Sumit Gulwani. 2022.
\newblock Synchromesh: Reliable code generation from pre-trained language
  models.
\newblock \emph{CoRR}, abs/2201.11227.

\bibitem[{Rabinovich et~al.(2017)Rabinovich, Stern, and Klein}]{RabinovichSK17}
Maxim Rabinovich, Mitchell Stern, and Dan Klein. 2017.
\newblock Abstract syntax networks for code generation and semantic parsing.
\newblock In \emph{{ACL} {(1)}}, pages 1139--1149.

\bibitem[{Raychev et~al.(2014)Raychev, Vechev, and Yahav}]{Raychev14}
Veselin Raychev, Martin~T. Vechev, and Eran Yahav. 2014.
\newblock Code completion with statistical language models.
\newblock In \emph{{PLDI}}, pages 419--428.

\bibitem[{Ren et~al.(2020)Ren, Guo, Lu, Zhou, Liu, Tang, Sundaresan, Zhou,
  Blanco, and Ma}]{CodeBLEU}
Shuo Ren, Daya Guo, Shuai Lu, Long Zhou, Shujie Liu, Duyu Tang, Neel
  Sundaresan, Ming Zhou, Ambrosio Blanco, and Shuai Ma. 2020.
\newblock Codebleu: a method for automatic evaluation of code synthesis.
\newblock \emph{CoRR}, abs/2009.10297.

\bibitem[{Salomaa et~al.(2001)Salomaa, Wood, and Yu}]{salomaa2001half}
Arto Salomaa, Derick Wood, and Sheng Yu. 2001.
\newblock \emph{A half-century of automata theory: celebration and
  inspiration}.
\newblock World scientific.

\bibitem[{Scholak et~al.(2021)Scholak, Schucher, and Bahdanau}]{PICARD}
Torsten Scholak, Nathan Schucher, and Dzmitry Bahdanau. 2021.
\newblock {PICARD:} parsing incrementally for constrained auto-regressive
  decoding from language models.
\newblock In \emph{{EMNLP} {(1)}}, pages 9895--9901. Association for
  Computational Linguistics.

\bibitem[{Sch{\"u}tzenberger(1963)}]{schutzenberger1963context}
Marcel~Paul Sch{\"u}tzenberger. 1963.
\newblock On context-free languages and push-down automata.
\newblock \emph{Information and control}, 6(3):246--264.

\bibitem[{Shen et~al.(2022)Shen, Zhu, Dong, Guo, Zhen, and
  Li}]{IndustryCodeGeneration}
Sijie Shen, Xiang Zhu, Yihong Dong, Qizhi Guo, Yankun Zhen, and Ge~Li. 2022.
\newblock Incorporating domain knowledge through task augmentation for
  front-end javascript code generation.
\newblock In \emph{{ESEC/SIGSOFT} {FSE}}, pages 1533--1543. {ACM}.

\bibitem[{Sun et~al.(2019)Sun, Zhu, Mou, Xiong, Li, and Zhang}]{SunZMXLZ19}
Zeyu Sun, Qihao Zhu, Lili Mou, Yingfei Xiong, Ge~Li, and Lu~Zhang. 2019.
\newblock A grammar-based structural {CNN} decoder for code generation.
\newblock In \emph{{AAAI}}, pages 7055--7062.

\bibitem[{Sun et~al.(2020)Sun, Zhu, Xiong, Sun, Mou, and Zhang}]{TreeGen}
Zeyu Sun, Qihao Zhu, Yingfei Xiong, Yican Sun, Lili Mou, and Lu~Zhang. 2020.
\newblock Treegen: {A} tree-based transformer architecture for code generation.
\newblock In \emph{{AAAI}}, pages 8984--8991.

\bibitem[{Vaswani et~al.(2017)Vaswani, Shazeer, Parmar, Uszkoreit, Jones,
  Gomez, Kaiser, and Polosukhin}]{Transformer}
Ashish Vaswani, Noam Shazeer, Niki Parmar, Jakob Uszkoreit, Llion Jones,
  Aidan~N. Gomez, Lukasz Kaiser, and Illia Polosukhin. 2017.
\newblock Attention is all you need.
\newblock In \emph{{NIPS}}, pages 5998--6008.

\bibitem[{Vinyals et~al.(2015)Vinyals, Fortunato, and Jaitly}]{PointerNetworks}
Oriol Vinyals, Meire Fortunato, and Navdeep Jaitly. 2015.
\newblock Pointer networks.
\newblock In \emph{{NIPS}}, pages 2692--2700.

\bibitem[{Wang et~al.(2022)Wang, Wang, Wan, Mi, Li, Zhou, Liu, Wu, Jiang, and
  Liu}]{Wang22}
Xin Wang, Yasheng Wang, Yao Wan, Fei Mi, Yitong Li, Pingyi Zhou, Jin Liu, Hao
  Wu, Xin Jiang, and Qun Liu. 2022.
\newblock Compilable neural code generation with compiler feedback.
\newblock In \emph{{ACL} (Findings)}, pages 9--19. Association for
  Computational Linguistics.

\bibitem[{Wang et~al.(2021)Wang, Wang, Joty, and teven C.~H.~Hoi}]{CodeT5}
Yue Wang, Weishi Wang, Shafiq~R. Joty, and teven C.~H.~Hoi. 2021.
\newblock Codet5: Identifier-aware unified pre-trained encoder-decoder models
  for code understanding and generation.
\newblock In \emph{{EMNLP} {(1)}}, pages 8696--8708.

\bibitem[{Wei et~al.(2019)Wei, Li, Xia, Fu, and Jin}]{Wei2019}
Bolin Wei, Ge~Li, Xin Xia, Zhiyi Fu, and Zhi Jin. 2019.
\newblock Code generation as a dual task of code summarization.
\newblock In \emph{NeurIPS}, pages 6559--6569.

\bibitem[{Xie et~al.(2021)Xie, Su, Ge, Li, Cui, Yao, and Wang}]{Xie21MLTranX}
Binbin Xie, Jinsong Su, Yubin Ge, Xiang Li, Jianwei Cui, Junfeng Yao, and Bin
  Wang. 2021.
\newblock Improving tree-structured decoder training for code generation via
  mutual learning.
\newblock In \emph{{AAAI}}, pages 14121--14128.

\bibitem[{Yin et~al.(2018)Yin, Deng, Chen, Vasilescu, and Neubig}]{YinDCVN08}
Pengcheng Yin, Bowen Deng, Edgar Chen, Bogdan Vasilescu, and Graham Neubig.
  2018.
\newblock Learning to mine aligned code and natural language pairs from stack
  overflow.
\newblock In \emph{{MSR}}, pages 476--486.

\bibitem[{Yin and Neubig(2017)}]{Yin17}
Pengcheng Yin and Graham Neubig. 2017.
\newblock A syntactic neural model for general-purpose code generation.
\newblock In \emph{{ACL} {(1)}}, pages 440--450.

\bibitem[{Yin and Neubig(2018)}]{TranX}
Pengcheng Yin and Graham Neubig. 2018.
\newblock {TRANX:} {A} transition-based neural abstract syntax parser for
  semantic parsing and code generation.
\newblock In \emph{{EMNLP} (Demonstration)}, pages 7--12.

\bibitem[{Yin and Neubig(2019)}]{Yin19Reranking}
Pengcheng Yin and Graham Neubig. 2019.
\newblock Reranking for neural semantic parsing.
\newblock In \emph{{ACL}}, pages 4553--4559.

\end{thebibliography}
\bibliographystyle{acl_natbib}

\end{document}